\documentclass[aps,prb,twocolumn,showpacs]{revtex4}
\usepackage{graphicx}
\usepackage{dcolumn}
\usepackage{amsmath}
\usepackage{amssymb}
\def\kF{k_{\rm F}}
\def\zF{z_{\rm F}}

\begin{document}
\title{Momentum distribution of
                         the uniform electron gas:\\ 
 improved parametrization and exact limits of the cumulant expansion}
\author{Paola Gori-Giorgi$^{1,2}$ and Paul Ziesche$^2$}
\affiliation{$^1$ INFM Center for
  Statistical Mechanics and Complexity, and
Dipartimento di Fisica, Universit\`a di Roma ``La Sapienza'', Piazzale 
A. Moro 2, I-00185 Rome, Italy\\
$^2$  Max-Planck-Institut f\"ur  Physik komplexer Systeme,
  N\"othnitzer Str. 38, D-01187 Dresden, Germany}
\date{\today}
\begin{abstract}
The momentum distribution of the unpolarized uniform electron gas in its
Fermi-liquid regime,
$n(k,r_s)$, with the momenta $k$ measured in units of the Fermi
wave number $\kF$ and
with the density parameter $r_s$, is constructed with the help 
of the convex Kulik function $G(x)$. It is assumed that
$n(0,r_s), \; n(1^\pm ,r_s)$, the on-top pair density 
$g(0,r_s)$ and the kinetic energy  $t(r_s)$
are known (respectively, from accurate calculations for $r_s=1,...,5$,
from the solution of the Overhauser model, and from Quantum Monte
Carlo calculations via the virial theorem). 
Information from the high- and the low-density limit, 
corresponding to the random-phase approximation and to the Wigner crystal
limit, is used. The result is an accurate parametrization
of $n(k,r_s)$, which fulfills most of the known exact constraints.
It is in agreement with the effective-potential calculations of
Takada and Yasuhara [Phys. Rev. B {\bf 44}, 7879 (1991)], is compatible 
with Quantum Monte Carlo data, and is valid in the density 
range $r_s\lesssim 12$.
The corresponding
cumulant expansions of the pair density and of the static structure factor are
discussed, and some exact limits are derived.
\end{abstract}
\pacs{71.10.Ca, 05.30.Fk}
\maketitle
\section{Introduction}
In solid state theory\cite{Ful} and quantum chemistry,\cite{Bar}
the phenomenon of electron correlation and some of its details 
are hidden in the reduced densities and reduced density 
matrices~\cite{Dav,Erd,Cole,Cio1} and their cumulants.\cite{Zie1,Zie2}
For the ground state of the uniform electron gas (jellium) of density
$\rho=3/(4\pi r_s^3)$ (in a.u.), these quantities are the pair
density $g(x,r_s)$ and the momentum distribution $n(k,r_s)$,
where $k$ is measured in units of the Fermi wave number $\kF=
(3\pi^2\rho)^{1/3}$ and $x$ is the
scaled interelectronic distance, $x=\kF r_{12}$.
\par
Besides its relevance in the understanding of many effects
in simple metals and semiconductors, the jellium model plays a
crucial role in providing  input quantities for approximate approaches to 
the many-electron problem of nonuniform density.
Different approximate schemes, in fact, often need different quantities
from jellium. As an example, density functional theory (DFT) uses the 
exchange-correlation energy of the uniform electron gas for the widely used
local density approximation (LDA). For building nonempirical 
beyond-LDA functionals for use in DFT, the pair density of jellium is often 
needed.\cite{GGA}

In such applications, the quantities from the uniform electron gas must
be available in the form of analytic expressions. Since the jellium model
is not exactly solvable, analytic expressions are built 
by interpolating between known exact limits, and by fitting the Quantum Monte
Carlo (QMC) data, when available. Relevant examples are (i) the correlation 
energy used in LDA implementation, built by using functional 
forms\cite{VWN,PZ,Per1}
which includes exact limits and interpolate the QMC data of Ceperley and
Alder;\cite{CA} (ii) the pair density, built by combining exact properties and
fitting to QMC data,\cite{Per2,Gor1} or by simply interpolating 
between exact limits;\cite{Gor3}
(iii) the static local field factor, parametrized by fitting\cite{Cor}
QMC data of the static response;\cite{Mor2} (iv) the dynamical local field
factors, built by using many known exact constraints.\cite{RA} 
All these parametrized
quantities are not strictly ``exact'', but are considered to be the closest
to the true quantities, in the sense that this is the best one can presently 
obtain, and in the sense that they are accurate enough for the purpose for 
which they are needed.

In recent years, there has been increasing interest in a particular
approach to the many-electron problem of nonuniform density, the so called
density matrix functional theory (DMFT), which uses the one-body
density matrix as basic variable.\cite{Cio1,Cio2,Cio3,CPZ} 
Building a ``local approximation''
for DMFT is not an easy task: in a first attempt,\cite{Yasuda} 
the momentum distribution $n(k,r_s)$
of jellium has been used as input.
Besides this important application, for which a reliable
parametrization of $n(k,r_s)$ is needed, there are other reasons to focus
on the momentum distribution of jellium.
The definitions of ``exchange'' and ``correlation'' in DMFT are
different from those in DFT and in Hartree-Fock-like approaches: in DMFT
the cumulant part\cite{Zie1,Zie2} of the pair density rises to be a
key quantity. An accurate parametrization of $n(k,r_s)$ at metallic 
densities allows to extract
the cumulant pair density of jellium, since
the whole pair density is available.\cite{Per2,Gor1,Gor3} 
The cumulant pair density can then
be compared with recent attempts to calculate
it in a high-density electron gas,\cite{Zie4a,Zie4b}
and can be diagonalized in terms of ``cumulant geminals''
(analog of ``Overhauser geminals'' for the pair density\cite{Ove,Gor2,Davo}). 
Also, the study of exact limiting
behaviors of the cumulant pair density is of great interest, since some
of these limits could be approximately valid in nonuniform systems.

The momentum distribution of the uniform electron gas is also useful
for the calculation of the exchange-correlation correction to 
Compton profiles computed within LDA-DFT.\cite{compton}
Note also that $n(k,r_s)$ determines the part of the local-field
factor beyond the random-phase approximation
(RPA) which takes into account the change in occupation numbers
in the Lindhard function [see Eq. (29) of Ref.~\onlinecite{RA}].\par

The qualitative behavior of $n(k,r_s)$ is the following.
It starts at $k=0$ with
a value $n(0,r_s)\lesssim 1$ and decreases with increasing $k$. For $k<1$
it is concave. Then in the Fermi-liquid regime 
at $k=1$ there is a finite jump (Fermi gap) from $n(1^-,r_s)$ to
a lower value
$n(1^+,r_s)=n(1^-,r_s)-\zF(r_s)$ with logarithmic slopes at both sides of
$k=1$. For $k>1$ (correlation tail) $n(k,r_s)$ is convex, and
vanishes for $k\to\infty$. 
For $r_s=0$ (ideal Fermi gas)
it is $n(k,0)=\theta(1-k)$, where $\theta(x)$ is the Heaviside step function.
Thus, the quasiparticle weight $\zF(r_s)$ starts with $\zF(0)=1$ and decreases
with increasing interaction strength $r_s$.
At large $r_s$ the electrons form a Wigner crystal with a smooth $n(k,r_s)$.
$r_s\ll 1$ and $r_s\gg 1$ are the weak- and strong-correlation limits,
respectively. For intermediate values of $r_s$ a non-Fermi liquid regime
may exist with $\zF=0$. In such case $n(k,r_s)$ would be continuous vs. $k$,
with a nonanalytical behavior at $k=1$.

In Ref.~\onlinecite{Zie3}, the idea of
using the convex Kulik function $G(x)$
to parametrize the two branches ($k<1$ and $k>1$) of $n(k,r_s)$ is
sketched.
The function $G(x)$ appeared in Kulik's~\cite{Kul}
RPA analysis of $n(k,r_s)$ near the Fermi
edge ($|1-k|<<1$), see Eq.~(\ref{eq_G(x)}). $G(x)$  
behaves as $c_0+c_1 x\ln x$ for small $x$ 
(see Appendix~\ref{app_RPA}
and Fig.~\ref{fig_kul}), which corresponds to the correct 
nonanalytic behavior of
$n(k,r_s)$ near the Fermi surface. 
So, supposing that the value at the centre, 
$n(0,r_s)$, and the values
at the Fermi edge, $n(1^-,r_s)$ and $n(1^+,r_s)$, are known,
it should be possible to represent $n(k,r_s)$ in terms of $G(x)$, 
with suitable prefactors and 
with suitable scaling (squeezing and stretching) of its argument.
In this way, $n(k,r_s)$ becomes a functional of $n_0(r_s)=n(0,r_s)$,
and of $n_{\pm}(r_s)=n(1^{\pm},r_s)$, and can be designed to yield
the proper normalization and the correct kinetic energy
$t(r_s)$, which follows from the total energy $\epsilon(r_s)=
t(r_s)+v(r_s)$ via the virial theorem.\cite{Mar1} 
In Ref.~\onlinecite{Zie3}, the input data from Takada and Yasuhara 
(TY)~\cite{Tak1,Tak2} for $n_0(r_s)$, $n_{\pm}(r_s)$ and
$t(r_s)$ at $r_s=1,...,5$ have been used, together with
the on-top pair density $g_0(r_s)=g(0,r_s)$ [which determines
the large-$k$ behavior of $n(k)$] from Ref.~\onlinecite{Gor2}. The result
is a field $n(k,r_s)$ for $r_s\in[1,6]$ which is correctly
concave for $k<1$, convex for $k>1$, and with a Fermi gap
$\zF(r_s)$ at $k=1$ decreasing with
growing $r_s$. The attempt to extend this procedure for $r_s\in[6,10]$
failed: $n(k<1,r_s)$ is no longer concave for $r_s\gtrsim 6$. \par

Here, an improved version of the parametrization of $n(k,r_s)$ in terms
of the Kulik function $G(x)$ is presented. Our parametrized momentum
distribution is
in good agreement with the TY values,\cite{Tak1,Tak2} and is valid in the 
range of densities $r_s\lesssim 12$. Previous 
parametrizations of $n(k,r_s)$~\cite{Ort1,Sen,Barb,Far} used the Quantum Monte
Carlo (QMC) data of Ref.~\onlinecite{Ort1} as an input. However, QMC data are 
presently only
available for $0.4\lesssim k \lesssim 0.9$ and $1.1 \lesssim k \lesssim 1.5$,
thus not providing information about $n(k,r_s)$ near the centre,
$k=0$, and at the Fermi edge, $k=1$. In these last regions, in fact,
different parametrizations of the same QMC data can be rather different 
from each other.\cite{Ort1,Sen,Barb}
Our construction of $n(k,r_s)$ uses
information from the effective-potential calculations of 
Takada and Yasuhara,\cite{Tak1,Tak2}
from the high- and low-density limits of $n(k,r_s)$, corresponding to
RPA and the Wigner crystal (WC) limit, and from accurate 
parametrizations
of $t(r_s)$~\cite{Per1} and of $g_0(r_s)$.\cite{Gor2} 
In the regions where QMC data are available,
our $n(k,r_s)$ is compatible with them. Also, with respect to previous 
works,\cite{Ort1,Sen,Barb,Far} the functional form used here 
satisfies more exact limits. In particular, the logarithmic
behavior at the Fermi edge is taken into account for the first
time.\cite{notaSav} 
Notice that it causes the logarithmic divergence of
$t(r_s\to 0)$.\cite{Cio4}\par
 
Using our $n(k,r_s)$,
the moments $\langle k^{\nu}\rangle$, the correlation entropy, and the
one-body reduced density matrix $f(x,r_s)$ are evaluated. 
The latter appears
in the cumulant partitioning of the pair density $g(x,r_s)$.
The static structure
factor $S(k,r_s)$, related to the pair density via Fourier transform, 
the particle-number fluctuations in fragments 
$\Delta N_{\Omega}(r_s)$, and the potential energy $v(r_s)$ are
discussed in terms of their cumulant partitioning, and some exact limits
are derived. Finally, by means of an accurate model for the 
spin-resolved pair density,\cite{Gor1} the cumulant part 
of $g(x,r_s)$ is extracted.\par

 The paper is organized as follows. In Sec.~\ref{sec_sumrules}, the
known sum rules and limiting cases for $n(k,r_s)$ are reported, and
they are used in Sec.~\ref{sec_parametriz.} to build up
our parametrization of the momentum distribution via the Kulik
function.
Sec.~\ref{sec_moments} is devoted to the calculation and
discussion of different measures of the correlation strength, of the
one-matrix, and of the cumulant expansion of the pair-density.
In Sec.~\ref{sec_Sq}, we study the cumulant partitioning
of the static structure factor, of the density fluctuations and of
the potential energy. Conclusions and 
future developments are reported in the last Sec.~\ref{sec_conc}.
\begin{figure}
\includegraphics[width=6.5cm]{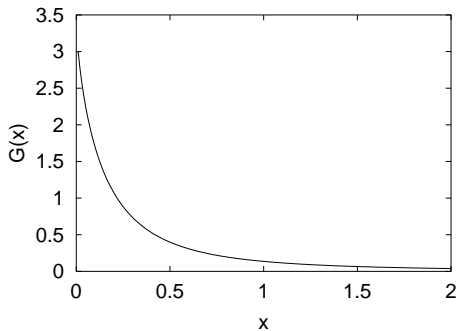}
 
\caption{The Kulik function $G(x)$ appearing in the RPA
analysis of $n(k,r_s)$, see Eq.~(\ref{eq_G(x)}). 
At the origin $G(x)$ has a finite value, $G(0)=3.353337$ [Eq.~(\ref{eq_Gof0})], and
a logarithmic-divergent slope, see Eq.~(\ref{eq_Gedge}).}
\label{fig_kul}
\end{figure}

\section{Sum rules and limiting cases} 
\label{sec_sumrules}
How is $n(k,r_s)$ defined? Starting from the many-body wave function
$\Psi(1,...,N)$, the one-body reduced density matrix (1-matrix for
short) results from the $N-1$ contraction
\begin{eqnarray}
& &\gamma(1|1')=\int\frac{d2...dN}{(N-1)!}\Psi(1,2,...,N)\Psi^*(1',2,...N),
\\ 
& & \int\frac{d1...dN}{N!}|\Psi(1,...,N)|^2=1, \nonumber
\end{eqnarray}
with the notation $1\equiv ({\underline r}_1,\sigma_1)$. For the uniform
electron gas, $\gamma(1|1')=\rho\,\delta_{\sigma_1\sigma_1'}\,f(\kF |
{\underline r}_1-{\underline r}_1'|,r_s)$ defines the dimensionless 1-matrix
$f(x,r_s)$. Then its Fourier transform is the momentum distribution
\begin{equation}
n(k,r_s)=\frac{\alpha^3}{2}\int_0^{\infty}dx^3\,\frac{\sin kx}{kx}\,f(x,r_s),
\label{eq_defn}
\end{equation}
where $dx^3=3/4\pi\,d^3x =3x^2dx$, and $\alpha=(4/9\pi)^{1/3}$.
$n(k,r_s)$ can be calculated using perturbation theory, directly with Green's
functions,\cite{Kul} see Figs.~1a and~1b of Ref.~\onlinecite{Zie4a}, or
via the Hellmann-Feynman 
theorem\cite{theorem} as the energy-derivative
$n(k,r_s)=\delta E/\delta\varepsilon_k$,
supposed $E$ is (perturbatively) known as a functional of $\varepsilon_k
=\hbar k^2/2m$ and $v_q=4\pi e^2/q^2$.\cite{Dan} Perturbative methods
only work for high-densities, $r_s\ll 1$. At metallic and lower
densities, other techniques must be used, namely, the effective-potential
method,\cite{Tak1} which combines
perturbation theory (Green's functions) with the Fermi-hypernetted chain
approach, and the QMC simulations.\cite{Ort1,Mor} A more
complete list of references concerning calculations and parametrizations
can be found in Ref.~\onlinecite{Zie3}.\par

$n(k,r_s)$ has to satisfy the condition $0<n(k,r_s)<1$ (which
guarantees the ensemble $N$-representability of the 1-matrix), and the
sum rules ($k$ in units of $\kF $, and energies in ryd)
\begin{eqnarray}
\int_0^{\infty}dk^3n(k,r_s)& = & 1, \label{eq_normal} \\
\frac{1}{(\alpha r_s)^2}\int_0^{\infty}dk^3 n(k,r_s)\,k^2 & = & t(r_s),
\label{eq_kinetic}
\end{eqnarray}
where $t(r_s)$ can be written as
the sum of the kinetic energy of the free Fermi gas, $3/5\,(\alpha r_s)^{-2}$,
and of the kinetic energy of correlation, $t_{\rm corr}(r_s)$. 
For $r_s\ll 1$, $t_{\rm corr}(r_s)$ is known from RPA and
from the lowest-order exchange diagram beyond it;\cite{Mac,Gell,Ons} for
a summary see Eq.~(3.25) and Figs.~1a and~1b of Ref.~\onlinecite{Zie4a}.
At larger $r_s$, $t_{\rm corr}(r_s)$ 
can be obtained via the virial theorem~\cite{Mar1}
from parametrized QMC correlation energies.\cite{Per1}\par
The large-$k$ behavior of $n(k,r_s)$ is determined by the kinks in the
many-body wave-function which occur whenever two electrons are 
at contact or ``on-top'' (coalescing cusp properties),\cite{Kim1,Yas}
\begin{eqnarray}
n(k\to\infty,r_s) & = &
\frac{C(r_s)}{k^8}+O\left(\frac{1}{k^{10}}\right),\nonumber \\
C(r_s) & = & \frac{8}{9\pi^2}(\alpha\,r_s)^2g_0(r_s), 
\label{eq_nlargek}
\end{eqnarray}
where $g_0(r_s)=g(0,r_s)$ is the on-top value of the pair-distribution
function. In the $r_s\to 0$ limit, $g_0(r_s)$ can be obtained from
perturbation theory,\cite{Gel,Kim2} and at larger $r_s$ it has
been calculated by solving an effective two-body Schr\"odinger 
equation.\cite{Gor2,notaRPA}\par

In a normal Fermi liquid,\cite{Pin} the momentum distribution has 
a discontinuity and infinite slopes~\cite{Sar} at the Fermi edge, $k=1$,
\begin{eqnarray}
n(k\to 1^-,r_s) & = & n_-(r_s)-A(r_s)(1-k)\ln(1-k)+\nonumber \\
& & O(1-k),
\label{eq_lim1-} \\
n(k\to 1^+,r_s) & = & n_+(r_s)+A(r_s)(k-1)\ln(k-1)+\nonumber \\
& & O(k-1).
\label{eq_lim1+} 
\end{eqnarray}
In the following, $A(r_s)$ is referred to as the Fermi edge coefficient.
In the small-$r_s$ limit, $n_\pm(r_s)$ and $A(r_s)$ are known
from RPA
(see Appendix~\ref{app_RPA}). In the low-density or WC limit,
the Fermi gap disappers [$n_-(r_s)\to n_+(r_s)$], the infinite
slopes at the Fermi edge may also vanish [$A(r_s)\to 0$].\par
Near the centre, $k\to 0$, $n(k,r_s)$ should behave 
quadratically,\cite{Sen,Barb,Far,Mor}
\begin{equation}
n(k\to 0,r_s)=n_0(r_s)+B(r_s)k^2+O(k^4).
\label{eq_nk0}
\end{equation}
A simple argument in favour of Eq.~(\ref{eq_nk0}) is that it holds both
in the high- and in the low-density limit (see Appendix~\ref{app_RPA} 
and~\ref{app_Wigner}).\par
When $r_s\to 0$, exact results for $n(k,r_s)$ are known by means of 
RPA~\cite{Dan,Kul}
(Appendix~\ref{app_RPA}). In the RPA treatment, the Kulik 
function $G(x)$ appears,\cite{Kul} see 
Eqs.~(\ref{eq_RPAnear1})-(\ref{eq_R(u)}) and Fig.~\ref{fig_kul}.
$G(x)$ will be used in the next section to build up a parametrized
$n(k,r_s)$ which satifies Eqs.~(\ref{eq_normal})-(\ref{eq_nk0}).\par
In the low-density or strongly correlated limit, $r_s\to\infty$,
the electron gas undergoes Wigner crystallization (see, e.g., 
Refs.~\onlinecite{Cepnew,Ort2}). A simple model for the 
momentum distribution
in such regime is reported in Appendix~\ref{app_Wigner}.

\section{Improved parametrization of $n(k,r_s)$}
\label{sec_parametriz.}
The momentum distribution in terms of the Kulik function
$G(x)$ of Fig.~\ref{fig_kul} is parametrized as follows. 
For $k<1$ we use the ansatz
\begin{equation}
n_<(k,r_s)  =  n_0-\frac{[n_0-n_-]}{G(0)}\,G[x_<(k,r_s)], 
\label{eq_nmin1}
\end{equation}
while for $k>1$ we use
\begin{equation}
n_>(k,r_s)  =  \frac{n_+}{G(0)}\,G[x_>(k,r_s)],
\label{eq_nmag1}
\end{equation}
with $x_<(k,r_s)$ and $x_>(k,r_s)$ equal to
\begin{eqnarray}
x_<(k,r_s) & = & a\frac{\alpha r_s}{2\pi^2}\frac{G(0)}{[n_0
-n_-]}\frac{(1-k)}{\sqrt{4\alpha r_s/\pi}} + \nonumber \\
& & 
b\frac{\pi^2}{\alpha r_s}\sqrt{\frac{\pi}{3}\frac{(1-\ln 2)}
{F''(0)}\frac{[n_0-n_-]}{G(0)}}\frac{(1-k)^2}{k},
\label{eq_xminor} \\
x_>(k,r_s) & = & a\frac{\alpha r_s}{2\pi^2}\frac{G(0)}{n_+}
\frac{(k-1)}{\sqrt{4\alpha r_s/\pi}}+\nonumber \\
& & \sqrt{\frac{3\pi(1-\ln 2)}{g_0}\frac{n_+}{G(0)}}\frac{\pi}{4\alpha
r_s}\,(k-1)^4. \label{eq_xmaggiore}
\end{eqnarray}
Here $F''(0)=17.968746$ [see Appendix~\ref{app_RPA}, 
Eq.~(\ref{eq_Fsmallk})], and the $r_s$
dependence of $a$, $b$, $n_0$, $n_{\pm}$ and $g_0$ is not explictly shown
for shortness. These constructions are such that $n_<(k)\to n_0,n_-$
for $k\to 0,1^-$, respectively, and $n_>(k)\to n_+,0$ for 
$k\to 1^+,\infty$, respectively. The behavior of the Kulik
function for small and large arguments (see Appendix~\ref{app_RPA})
ensures the exact asymptotic expansion of 
Eqs.~(\ref{eq_nlargek})-(\ref{eq_nk0})
near the centre, near the Fermi surface, and for large $k$.
\par
The parameter $a(r_s)$ determines the Fermi edge coefficient 
$A(r_s)$ of the $|1-k|\ln|1-k|$
term at the Fermi surface,
\begin{eqnarray}
n(k\to 1^\pm,r_s) &  = & n_{\pm}(r_s)\pm a(r_s)\left(\frac{\alpha r_s}
{\pi}\right)^{1/2}\frac{1}{4}\left(\frac{\pi}{4}+\sqrt{3}\right)\times
\nonumber \\
& & |1-k|\ln|1-k|+O(|1-k|).
\end{eqnarray}
The parameter $b(r_s)$ determines the curvature $B(r_s)$ 
of Eq.~(\ref{eq_nk0}) at the centre, $k=0$,
\begin{equation}
n(k\to 0,r_s) = n_0(r_s)-\frac{\pi^4}{\alpha^2}\frac{F''(0)}{2}
\left[\frac{r_s}{b(r_s)}\right]^2k^2+O(k^4).
\end{equation}
For small $r_s$ (RPA - Appendix~\ref{app_RPA}) it is $a(r_s\to 0)=1$ and
$b(r_s\to 0)=1$.
\par
In the preliminary version of Ref.~\onlinecite{Zie3}, another (but similar)
ansatz was introduced, and it was chosen $b(r_s)=1$. Two different 
functions, $a_<(r_s)$ and $a_>(r_s)$, for the coefficient of
$|1-k|\ln|1-k|$ at $k=1^-$ and $k=1^+$  were fixed by the sum
rules of Eqs.~(\ref{eq_normal}) and~(\ref{eq_kinetic}) (with
$t_{\rm corr}(r_s)$ from Ref.~\onlinecite{Per1}). The values $n_0(r_s)$ and
$n_{\pm}(r_s)$ were taken from the TY data (available for
$r_s=1,...,5$). The on-top value $g_0(r_s)$ was taken from 
Ref.~\onlinecite{Gor2}.\par
In our improved ansatz of Eqs.~(\ref{eq_nmin1})-(\ref{eq_xmaggiore}),
we set $a_<(r_s)=a_>(r_s)=a(r_s)$ (in agreement with Ref.~\onlinecite{Sar};
also Figs.~7 and~8 of Ref.~\onlinecite{Zie3} confirm this), 
and we use again $t_{\rm corr}(r_s)$ from Ref.~\onlinecite{Per1} and $g_0(r_s)$ from
Ref.~\onlinecite{Gor2}. Since we want to extend our results in the
density range $6\le r_s\le 10$, where there are no data available
for $n_0(r_s)$ and $n_{\pm}(r_s)$, we extract information from the extreme
low-density limit (Wigner crystal - see Appendix~\ref{app_Wigner}), by
following an oversimplified version of the idea
presented in Ref.~\onlinecite{Sei}.\par
\begin{figure}
\includegraphics[width=7cm]{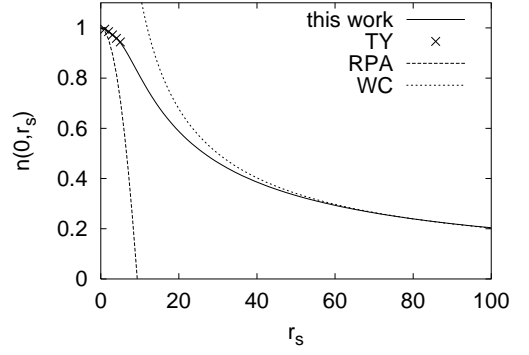}
 
\caption{Parametrized $n(0,r_s)$ (solid line), compared to the
Takada-Yasuhara (TY) values.\cite{Tak1,Tak2} The high-density 
or RPA limit and
the Wigner crystal (WC) limit are also shown.}
\label{fig_n0}
\end{figure}
\begin{figure}
\includegraphics[width=7cm]{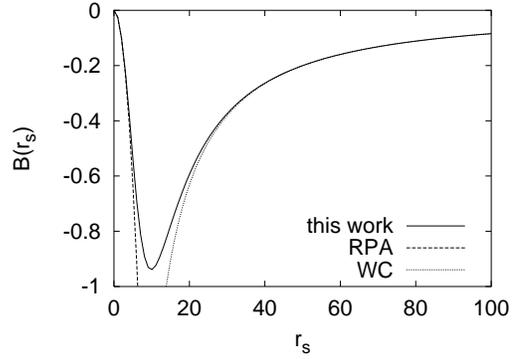}
 
\caption{Parametrized coefficient of the $k^2$ term
near the centre ($k\to 0$), $B(r_s)=-\frac{\pi^4}{\alpha^2}\frac{F''(0)}{2}
\left[\frac{r_s}{b(r_s)}\right]^2$. The high-density or RPA
result, $b(r_s\to 0)=1$, and the Wigner crystal (WC) limit are also shown.}
\label{fig_b}
\end{figure}
We first build $n_0(r_s)$ by using a functional form which
recovers the exact high-density limit, includes the Wigner crystal
behavior as $r_s \to \infty$, and has some free parameters to be
fitted to the TY data. The result is reported in Fig.~\ref{fig_n0}, together
with the high- and low-density curves. It is given by
\begin{equation}
n_0(r_s)=\frac{1+t_1\,r_s^2+t_2\,r_s^{5/2}}{1+t_3\,r_s^2+t_4\,r_s^{13/4}},
\label{eq_n0}
\end{equation}
with $t_1=0.003438169$, $t_2=0.00725313666$, $t_3=0.014900367$,
$t_4=0.00113244364$ ($t_1-t_3$ agrees with the RPA value
$-(\frac{\alpha}{\pi^2})^2\, 4.1123=-0.01146$). 
We then build the parameter $b(r_s)$ by
a simple interpolation between the high- and low-density limit
of the curvature at the centre (see Fig.~\ref{fig_b}). The result
is
\begin{equation}
b(r_s)=\left(1+0.0009376925\, r_s^{13/4}\right)^{1/2}.
\label{eq_b}
\end{equation}
\begin{figure}
\includegraphics[width=7cm]{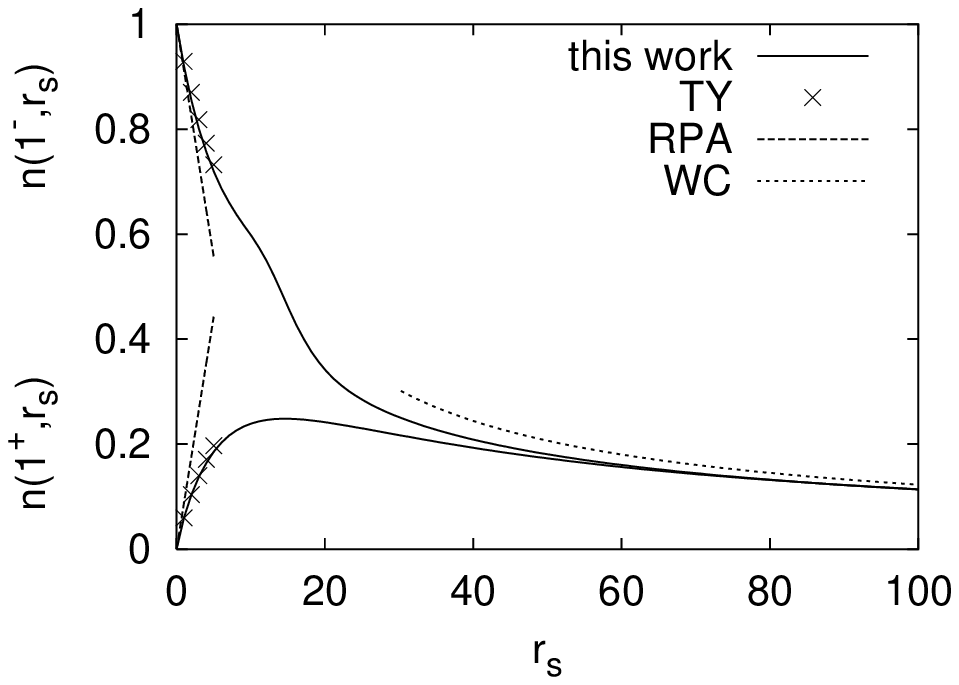}
\includegraphics[width=7cm]{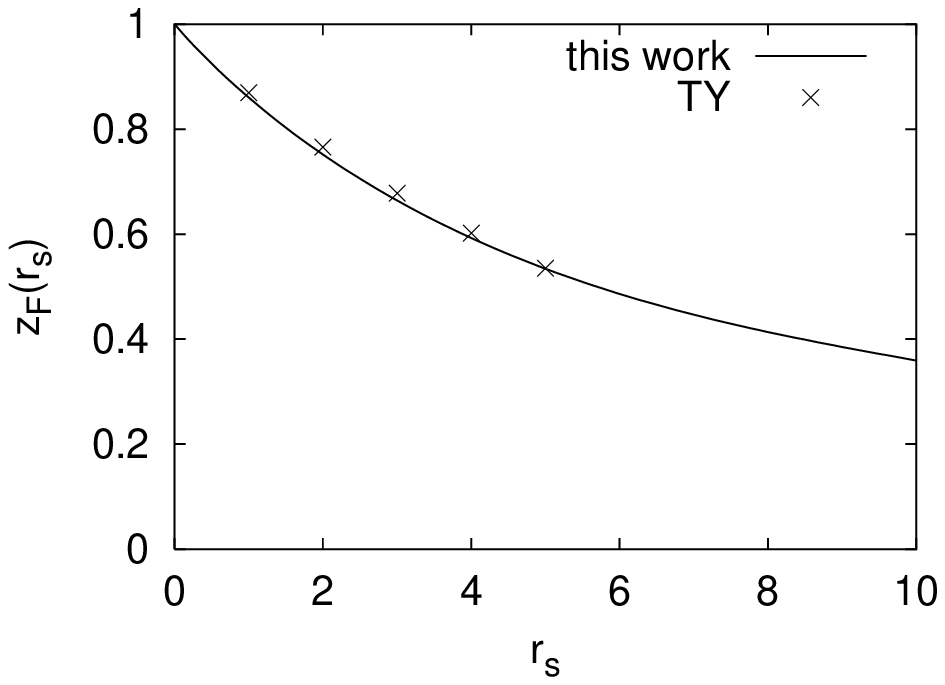}
 
\caption{Upper panel: parametrized $n(1^\pm,r_s)$ (solid lines), compared to
the Takada-Yasuhara (TY) values.\cite{Tak1,Tak2} The high-density or
RPA limit and
the Wigner crystal (WC) limit are also shown. Lower panel:
value of the Fermi gap $\zF(r_s)=n(1^-,r_s)-n(1^+,r_s)$ as a 
function of $r_s$;
the present parametrization is compared with the TY results.\cite{Tak1,Tak2}}
\label{fig_zF}
\end{figure}
\begin{figure}
\includegraphics[width=7cm]{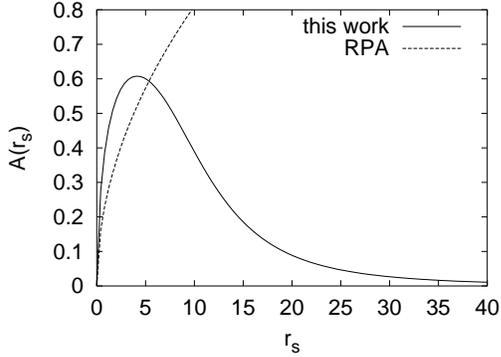}
 
\caption{Parametrized coefficient of the infinite
slope at the Fermi edge, $A(r_s)=a(r_s)\left(\frac{\alpha r_s}{\pi}\right)
^{1/2}0.63$. The present result is
compared with the RPA value, $a(r_s\to 0)=1$. 
Notice Ref.~\protect\onlinecite{notaSav}.}

\label{fig_a}
\end{figure}
\begin{figure}
\includegraphics[width=7cm]{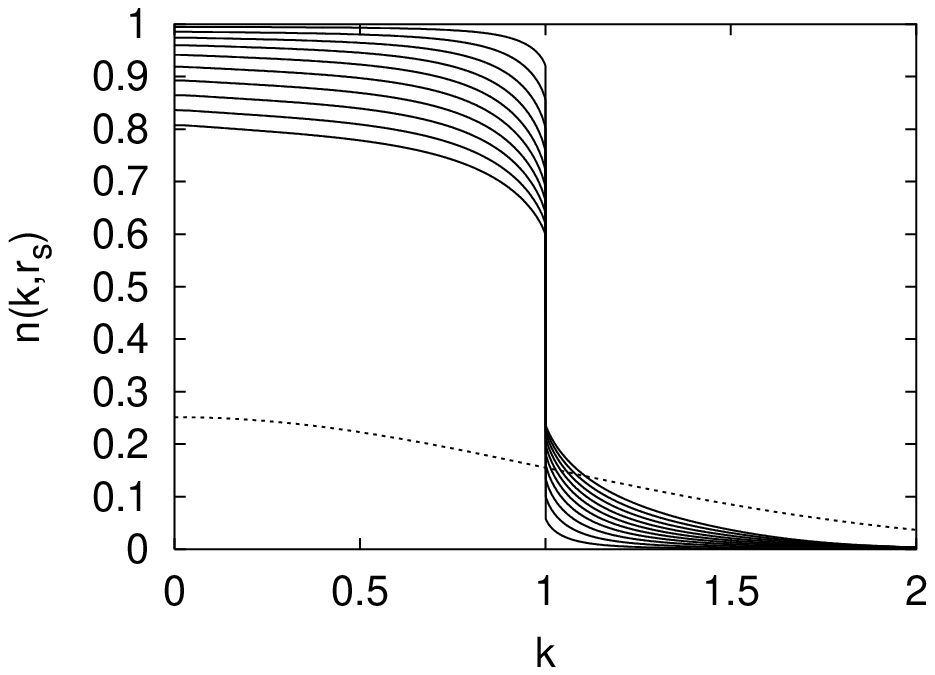}
\includegraphics[width=7cm]{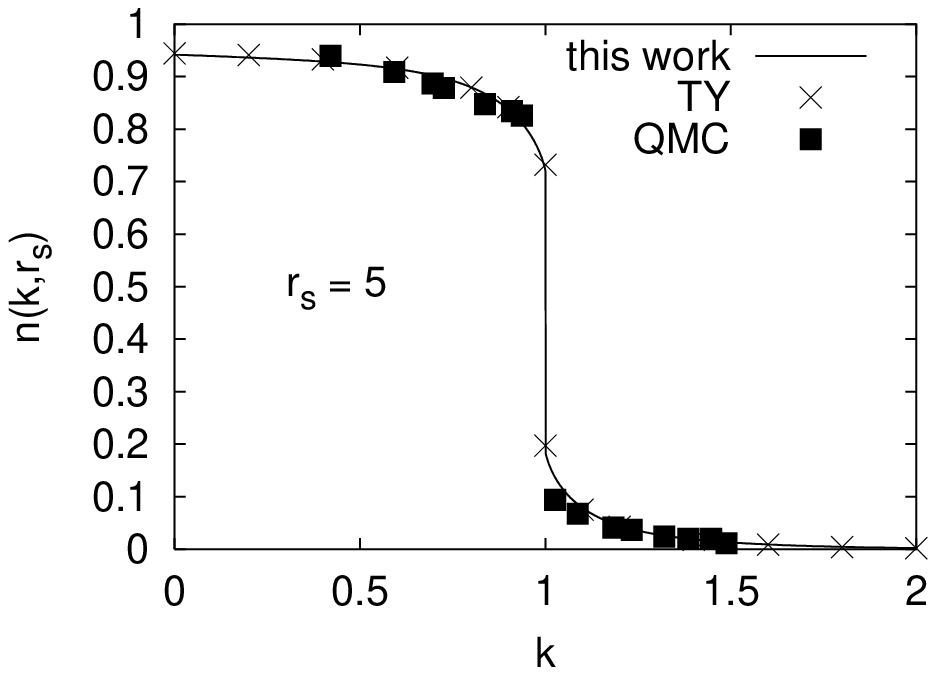}
\caption{Upper panel: momentum distribution calculated with 
Eqs.~(\ref{eq_nmin1})-(\ref{eq_xmaggiore})
for $r_s=1,2,...,10$ (solid lines). In the Wigner limit,
$n(k,r_s\gg 1)$ is calculated with Eq.~(\ref{A2.1}) (dashed line, corresponding
to $r_s=75$). Lower panel: comparison of the present work with
the Takada-Yasuhara (TY)~\cite{Tak1,Tak2} momentum distribution and the QMC 
calculation of Ref.~\protect\onlinecite{Ort1} for $r_s=5$.}
\label{fig_allrs}
\end{figure}
Finally, the values at the Fermi edge, $n_{\pm}(r_s)$, and 
the coefficient of the infinite slope at the Fermi edge, $a(r_s)$,
are obtained by fitting the TY values for $n_{\pm}(r_s)$
while imposing the normalization and the kinetic
energy~\cite{Per1} sum rules of Eqs.~(\ref{eq_normal})
and~(\ref{eq_kinetic}).
The results are parametrized with the inclusion of
the high- and low- density limits, and are equal to
\begin{equation}
n_- (r_s)= \frac{1+v_1\,r_s+v_2\, r_s^2+v_3\,r_s^3}{1+v_4 r_s+v_5 r_s^2+
v_6\,r_s^3+v_7\,r_s^{15/4}},  
\label{eq_n1-}
\end{equation}
with $v_1= -0.0679793$, $v_2= -0.00102846$, $v_3=0.000189111$,
$v_4=0.0205397$, $v_5= -0.0086838$, $v_6=6.87109\cdot 10^{-5}$,
$v_7=4.868047\cdot 10^{-5}$ ($v_1-v_4$ agrees with the RPA value
$-\frac{\alpha}{2\pi^2}\, 3.3533=-0.088519$), and
\begin{equation}
n_+(r_s)=\frac{q_1\,r_s}{1+q_2\,r_s^{1/2}+q_3\,r_s^{7/4}}, 
\label{eq_n1+}
\end{equation}
with $q_1=0.088519$ (from RPA), $q_2=0.45$, $q_3=0.022786335$;
\begin{equation}
a(r_s)=\frac{1+p_1\,r_s^{1/4}+p_2\,r_s^{1/2}}{1+p_3\,r_s^{1/4}+p_4\,
r_s^{1/2}+p_5\,r_s+p_6\,r_s^6},
\label{eq_a}
\end{equation}
with $p_1=-78.8682$, $p_2=-0.0989941$, $p_3=-68.5997$,
$p_4=38.1159$, $p_5=-17.6829$, $p_6=-0.01136759$. Our
parametrized $n(k,r_s)$ breaks down at $r_s\gtrsim 12$ [in the
density range $12\lesssim r_s \lesssim 16$, $n(k>1,r_s)$ is no longer convex,
and for $r_s\gtrsim 16$ the unphysical result $n_-<n_+$ is obtained
when the sum rules of Eqs.~(\ref{eq_normal}) and~(\ref{eq_kinetic}) 
are imposed].
\par
In the upper panel of Fig.~\ref{fig_zF}, we show the functions
$n_{\pm}(r_s)$, together with the TY values, and the high and 
low-density limits (here the  $r_s\to\infty$ limit is considered
to be the inflexion point of the WC momentum distribution,
see Eq.~(\ref{A2.4})). 
As said, our model is only valid for
$r_s \lesssim 12$, so that $n_-(r_s)$ and $n_+(r_s)$ at densities lower
than $r_s=12$ are no more obtained from the constraints of 
Eqs.~(\ref{eq_normal}) and~(\ref{eq_kinetic}). Thus, the strange behavior
of $n_-(r_s)$ at $r_s\sim 16$ does not affect our results.
 Also, the scheme presented here for the
transition between the metallic and the extreme low-density region is
oversimplified and must not be regarded as rigorous or reliable: we did not
take into account the transition to the partially polarized 
electron gas (which
affects the $r_s\gtrsim 50$ densities~\cite{Cepnew}), 
as well as many other features. However, our results seem to be
reliable in the relevant density range $r_s\lesssim 12$, and the 
simple picture of the upper panel of Fig.~\ref{fig_zF} is only a 
``na\"ive suggestion''. In the
lower panel of Fig.~\ref{fig_zF}, we compare our parametrized $\zF(r_s)$
with the TY results. In Fig.~\ref{fig_a}, we report the
$r_s$ dependence of the Fermi edge coefficient $A(r_s)$.
 Finally, in Fig.~\ref{fig_allrs},
we present in the upper panel our parametrized $n(k,r_s)$ for 
$1\le r_s\le 10$, and in the lower panel we compare our result with
the TY $n(k,r_s)$ and with the QMC data of Ref.~\onlinecite{Ort1} for 
$r_s = 5$.

\section{Moments, correlation entropy, 1-matrix, and cumulant expansion}
\label{sec_moments}
\begin{figure}
\includegraphics[width=7cm]{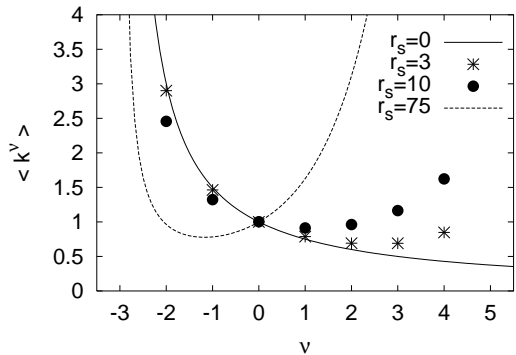}
\caption{The moments $\langle k^{\nu}\rangle$ of $n(k,r_s)$ for
$r_s=3$ and $r_s=10$. The corresponding results for the noninteracting
gas ($r_s=0$) and for the Wigner crystal at $r_s=75$ are also reported.}
\label{fig_momenta}
\end{figure}

With the now available momentum distribution $n(k,r_s)$, its 
moments
\begin{equation}\label{4.1}
\langle k^\nu\rangle=\int_0^\infty dk^3\; n(k,r_s) k^\nu,
\end{equation}
can be evaluated in addition to the normalization for $\nu=0$ and
the kinetic energy for $\nu=2$ [Eqs.~(\ref{eq_normal}) 
and~(\ref{eq_kinetic})]. They are shown in
Fig.~\ref{fig_momenta} for $r_s=3$ and 10, together with
the $r_s=0$ (ideal Fermi gas) and with the WC
result ($r_s=75$). The expression
\begin{equation}\label{4.2}
(\Delta t)^2 = \frac{1}{(\alpha r_s)^4}[\langle k^4\rangle-\langle k^2
\rangle^2]
\end{equation}  
(measured in ${\rm ryd}^2$) describes the fluctuation of the 
kinetic energy. The moments $\langle k^2\rangle $ and 
$\langle k^4\rangle$ determine the
small-$x$ behavior of the 1-matrix, see Eq.~(\ref{4.6}).\par
\begin{figure}
\includegraphics[width=7cm]{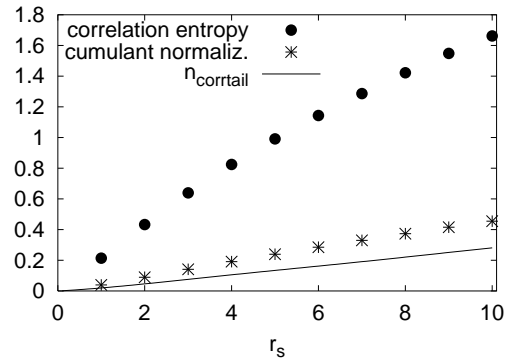}
\caption{The particle-hole symmetric correlation entropy
[Eq.~(\ref{4.3})], the normalization of the cumulant pair density
[r.h.s.~of Eq.~(\ref{5.3})], and the correlation tail normalization 
$n_{\rm corrtail}(r_s)$ [Eq.~(\ref{4.4})] as a function of $r_s$.}
\label{fig_corrent}
\end{figure}

In Refs.~\onlinecite{Zie6,Zie7}, the entropy-like expression 
$s(r_s)=-\langle \ln n(k,r_s)\rangle$ as a 
function of the interaction strength $r_s$ has been used as a 
measure of the correlation strength.\cite{notas} Here the expression
\begin{eqnarray}
s_{\rm ph}(r_s) & = & \int dk^3\; (-1)\{ n(k,r_s)\ln n(k,r_s)+
\nonumber \\ & & [1-n(k,r_s)]\ln [1-n(k,r_s)]\}
\label{4.3} 
\end{eqnarray} 
is introduced as an alternative with the understanding that 
$n(k,r_s)$ and $1-n(k,r_s)$ are the probabilities for the 
momentum state $\underline k$ to be occupied (with spin up and spin down)
and empty, 
respectively. The entropy of this probability `distribution' is
just the integrand of Eq.~(\ref{4.3}), and $s_{\rm ph}(r_s)$ is the sum of
all these entropies. Notice its invariance under the exchange 
$n(k,r_s)\leftrightarrow 1-n(k,r_s)$, which is referred to as 
particle-hole symmetry in the Reduced-Density-Matrix community. 
This symmetry is an
intrinsic property of the correlation energy as a functional of 
the 1-matrix.\cite{Rus} $s_{\rm ph}(r_s)$ is plotted in 
Fig.~\ref{fig_corrent}. 
Another measure of the correlation strength
is the correlation-tail normalization
\begin{equation}\label{4.4}
n_{\rm corrtail}(r_s)=\int_1^\infty dk^3\; n(k,r_s),
\end{equation}
also reported in Fig.~\ref{fig_corrent}. For large $r_s$,
the Fermi edge disappears, $z_{\rm F}(r_s)=0$, and also 
any relict of it, $A(r_s)=0$,  then the inflexion point of 
$n(k,r_s)$ vs. $k$ may serve in Eq.~(\ref{4.4}) as the lower 
limit. 

\begin{figure}
\includegraphics[width=7cm]{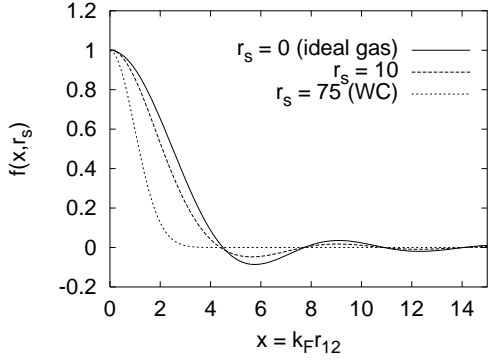}
\caption{1-matrix $f(x,r_s)$ for $r_s=0$ (ideal Fermi gas), for
$r_s = 10$ (present model), and for $r_s=75$ [Wigner crystal (WC) limit,
Eq.~(\ref{A2.1})].}
\label{fig_onema}
\end{figure}

\begin{figure}
\includegraphics[width=7cm]{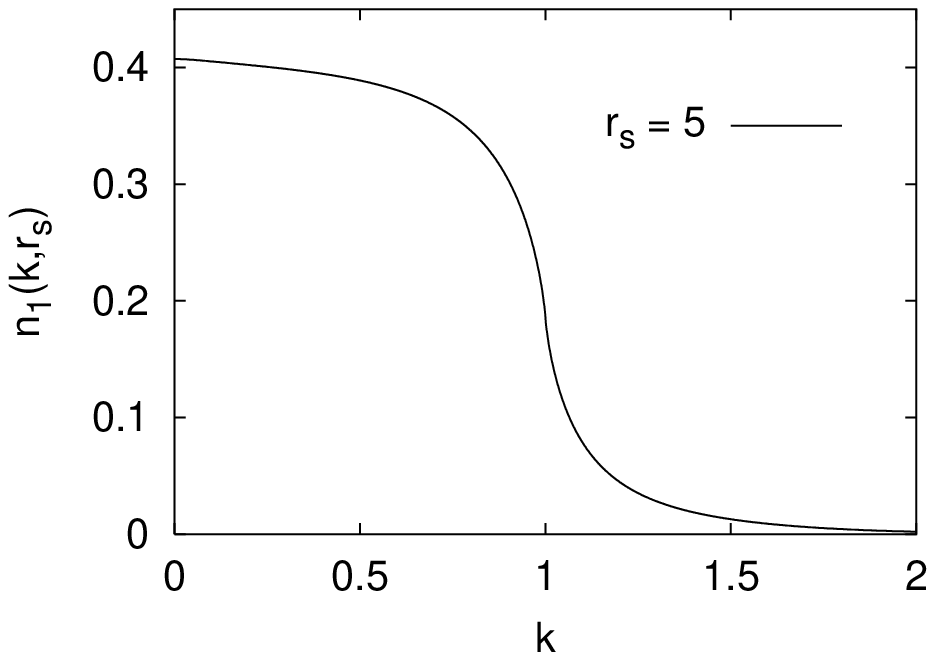}
\includegraphics[width=7cm]{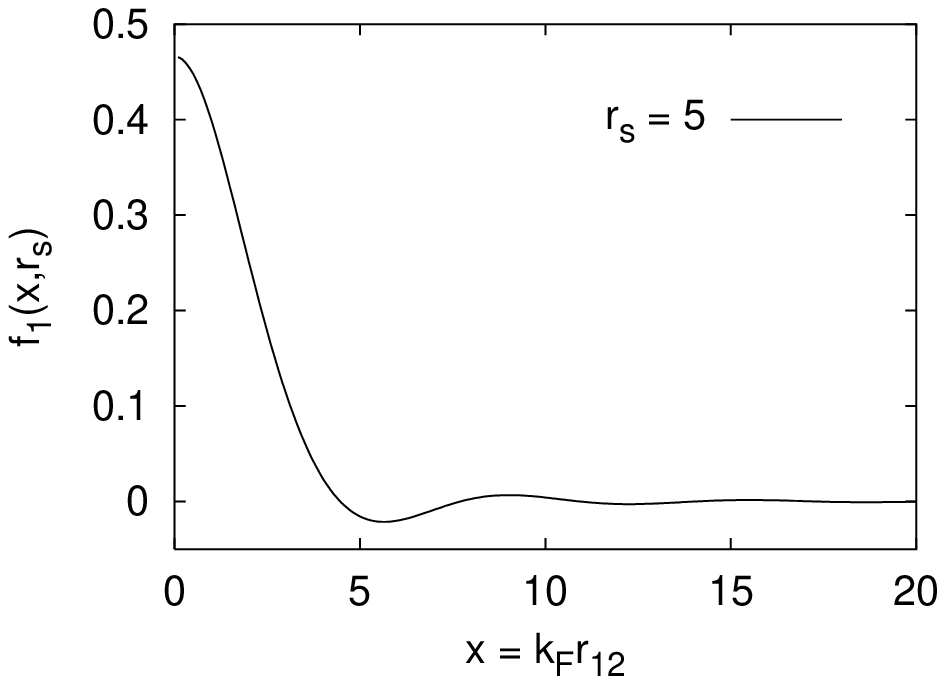}
 \caption{Continuous part of $n(k,r_s)$, $n_1(k,r_s)=
n(k,r_s)-\zF(r_s)\theta(1-k)$ (upper panel), and the
corresponding 1-matrix $f_1(x,r_s)$ (lower panel). The oscillations
of $f_1(x)$ are due to the infinite slope of $n_1(k)$ at $k=1$.}
\label{fig_n1f1}
\end{figure}

With $n(k,r_s)$ also the (dimensionless) 1-matrix 
\begin{equation}\label{4.5}
f(x,r_s)=\int_0^\infty dk^3\,\frac{\sin kx}{kx}\,n(k,r_s), \quad 
x=k_{\rm F}|\underline r-\underline r'|
\end{equation}
is available as the inverse of Eq.~(\ref{eq_defn}). 
It has the small-$x$ behavior
\begin{eqnarray}
f(x\ll 1,r_s)& = & 1-\frac{\langle k^2\rangle}{3!}\cdot x^2+
\frac{\langle k^4\rangle}{5!}\cdot x^4 \nonumber \\
& & -{1\over 5!}\frac{2}{9\pi}
(\alpha r_s)^2g_0(r_s)\cdot x^5 +O(x^6)
\label{4.6}
\end{eqnarray} 
and the large-$x$ asymptotics (Friedel oscillations with reduced
amplitudes)
\begin{eqnarray}
f(x\gg 1,r_s)& = & -3\,z_{\rm F}(r_s)\,\frac{\cos x}{x^2}+ 
\frac{3}{x^3}\,[z_{\rm F}(r_s)\sin x-\nonumber \\ 
& & \pi A(r_s)\cos x]+ O\left({1\over x^4}\right),
\label{4.7}
\end{eqnarray} 
see Appendix~\ref{app_diag}. 
$A(r_s)=a(r_s)\,(\frac{\alpha r_s}{\pi})^{1/2}0.63$ is the Fermi edge
coefficient,
the prefactor of the logarithmic term $(k-1)\ln |k-1|$ in 
$n(k\approx 1,r_s)$. The factor 0.63 is the Kulik number 7.91,
see Eq.~(\ref{eq_Gedge}), divided by $4\pi$. 
In the inverse Fourier transform~(\ref{eq_defn}), the oscillatory terms
of Eq.~(\ref{4.7}) do not 
affect the small-$k$ behavior of $n(k,r_s)$, because their 
average is zero. Since $n(k\ll 1,r_s)=n_0(r_s)+O(k^2)$, the 
large-$x$
behavior of the non-oscillatory $f(x,r_s)$ is $\propto 1/x^6$ or faster.
$f(x,r_s)$ is displayed in Fig.~\ref{fig_onema}. One may partition $n(k,r_s)$
and correspondingly $f(x,r_s)$ in the following way:
\begin{eqnarray}
n(k,r_s) & = & z_{\rm F}(r_s)\theta(1-k) +n_1(k,r_s), \nonumber \\
\label{4.8}
f(x,r_s) & = & 3\,z_{\rm F}(r_s)\,\frac{ j_1(x)}{x}+f_1(x,r_s),
\end{eqnarray}
where $j_1(x)=(\sin x-x\cos x)/x^2$. $n_1(k,r_s)$ is a 
continuous function with $n_1(1^-,r_s)=n_1(1^+,r_s)$ and 
an infinite slope at $k=1$. 
Figure~\ref{fig_n1f1} shows $n_1(k,r_s)$ and $f_1(x,r_s)$
for $r_s=5$. \par
The 1-matrix squared appears in the cumulant partitioning of the
pair density,
\begin{equation}\label{4.9}
g(x,r_s)=1-{1\over 2}|f(x,r_s)|^2-h(x,r_s),\quad 
x=k_{\rm F}r_{12}.
\end{equation}
This defines the cumulant pair density $h(x,r_s)$, which is the
diagonal of the cumulant 2-matrix 
$\chi(1|1',2|2')$. The 
spin-resolved version of Eq.~(\ref{4.9}) is
\begin{eqnarray}
& & g_{\uparrow\uparrow}(x,r_s)=1-|f(x,r_s)|^2-
h_{\uparrow\uparrow}(x,r_s),\nonumber \\
& & g_{\uparrow\downarrow}(x,r_s)=1- h_{\uparrow\downarrow}(x,r_s)
\label{4.10}
\end{eqnarray}
with $g(x,r_s)={1\over 2}[g_{\uparrow\uparrow}(x,r_s)+
g_{\uparrow\downarrow}(x,r_s)]$ and $h(x,r_s)={1\over 2}
[h_{\uparrow\uparrow}(x,r_s)+h_{\uparrow\downarrow}(x,r_s)]$.
 Notice that the (generalized exchange or) Fock term $|f(x,r_s)|^2$ appears 
only in the parallel-spin pair density and not in the 
antiparallel-spin pair density. $g_{\uparrow\uparrow}(x,r_s)$
describes the Fermi hole (due to both Pauli and Coulomb 
repulsion) with 
$g_{\uparrow\uparrow}(0,r_s)=h_{\uparrow\uparrow}(0,r_s)=0$ and 
$g_{\uparrow\downarrow}(x,r_s)$ describes the Coulomb hole 
(only due to the Coulomb repulsion) with $g(0,r_s)<1$. In 
addition to the above mentioned
correlation-strength indices, the quantities 
$h''_{\uparrow\uparrow}(0,r_s)$ measuring the on-top Fermi-hole
curvature, and $h_{\uparrow\downarrow}(0,r_s)$ measuring the 
on-top Coulomb hole are other ones.\par

From Eq.~(\ref{4.7}) it follows
\begin{eqnarray}
|f(x\gg 1,r_s)|^2 & = & \frac{9}{2}
\left [\frac{z_{\rm F}^2(r_s)}{x^4}+
\frac{2 \pi A(r_s) z_{\rm F}(r_s)}{x^5}\right](1+\cos 2x)
\nonumber \\ 
& & -
9\frac{ z_{\rm F}^2(r_s)}{x^5}\sin 2x+O\left(\frac{1}{x^6}\right).
\label{4.11}
\end{eqnarray}
If this is inserted into Eq.~(\ref{4.9}), then the 
non-oscillatory terms, $\frac{9}{2}\frac{\zF^2}{x^4}+
\frac{9\pi\zF A}{x^5}$, are canceled by the asymptotics of 
$h(x,r_s)$, which follow from the sum-rule properties of the 
static structure factor $S(q,r_s)$, see Sec.~\ref{sec_Sq} and 
Ref.~\onlinecite{Kim2}.
The nominator of the oscillating $1/x^5$ term can be written as
\begin{eqnarray*}
& & 9\zF{\sqrt {z_{\rm F}^2(r_s)+\pi^2A^2(r_s)}} \cos (2x+2x_0(r_s)), \\
& & \tan 2x_0(r_s)=\frac{z_{\rm F}(r_s)}{\pi A(r_s)}.
\end{eqnarray*}
The on-top properties of $g_{\uparrow\uparrow}(x\ll 1,r_s)$
and $g_{\uparrow\downarrow}(x\ll 1,r_s)$ are determined by the 
coalescing cusp theorems.\cite{Kim1,Kim3}  

\begin{figure}
\includegraphics[width=7cm]{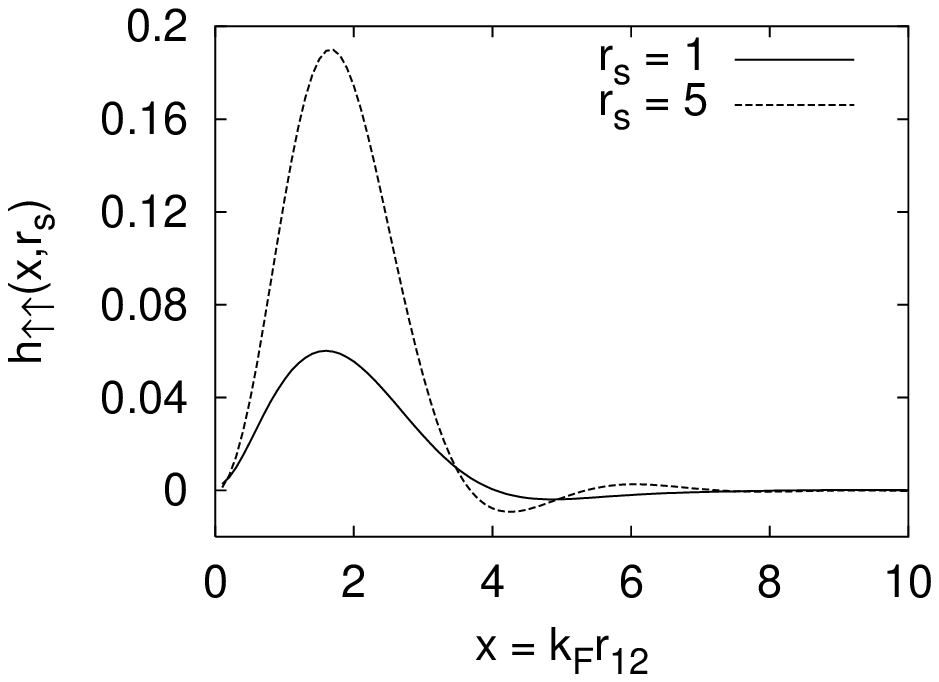}
\includegraphics[width=7cm]{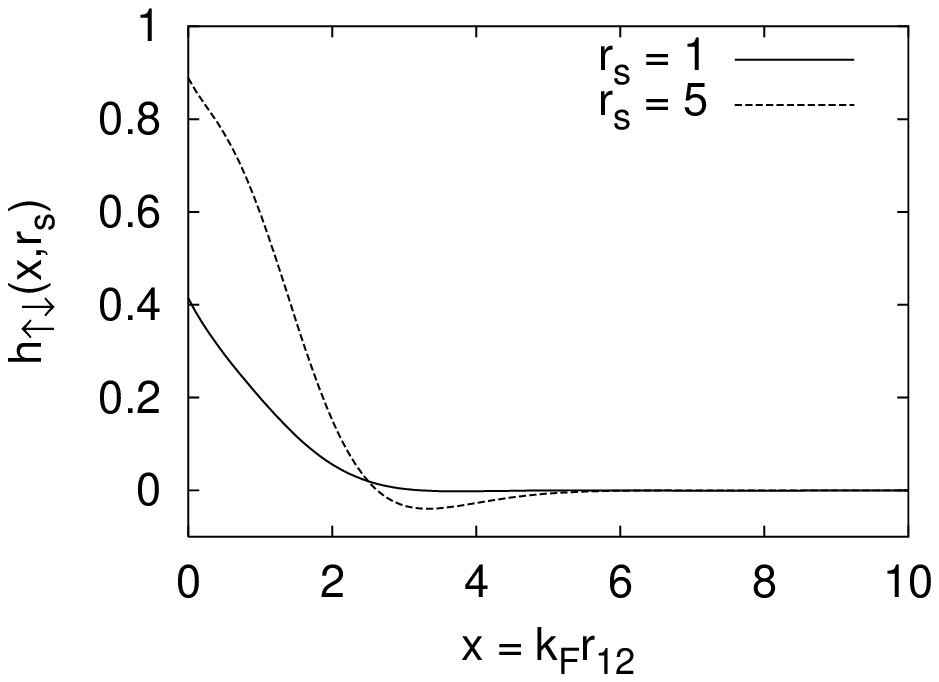}
\caption{Cumulant pair densities for parallel and antiparallel
spins, obtained by combining the present work with the results
of Ref.~\protect\onlinecite{Gor1}.}
\label{fig_h}
\end{figure}

With Eq.~(\ref{4.10}), with the spin-resolved pair densities of 
Ref.~\onlinecite{Gor1}, and with $f(x,r_s)$ of this paper, the 
resulting cumulant pair densities $h_{\uparrow\uparrow}(x,r_s)$ 
and $h_{\uparrow\downarrow}(x,r_s)$ are plotted in Fig.~\ref{fig_h}. 
For small $r_s$ ($\ll 1$), our results agree with those of Ref.~\onlinecite{Zie4a}. 

\section{Static structure factor, density fluctuations, and potential energy}
\label{sec_Sq}
The cumulant partitioning of Eq.~(\ref{4.9}) causes corresponding 
decompositions of all the quantities containing $1-g(x,r_s)$.
Such quantities are the static structure factor $S(q,r_s)$, the
fluctuation $\Delta N_{\Omega}(r_s)$ of the particle number in a
fragment $\Omega$, and the potential energy $v(r_s)$. \par

So, the static structure factor is given by 
\begin{equation}\label{5.1a}
S(q,r_s)=1-\frac{1}{2}{\tilde n}^2(q,r_s)-{\tilde h}(q,r_s)
\end{equation}
with the (generalized exchange or) Fock component
\begin{eqnarray}
& & {\tilde n}^2(q,r_s)  =  \alpha^3\int_0^\infty dx^3\; 
\frac{\sin qx}{qx}|f(x,r_s)|^2 = \nonumber \\
\label{5.1}
&  & \int_0^\infty dk^3\; n(k,r_s)\int_{-1}^{+1}d\zeta\,
n({\sqrt {k^2+q^2-2kq\zeta}},r_s)
\end{eqnarray}
(from which it follows that $\tilde{n}^2(q,r_s)$ has a discontinuity
in its second derivative at $q=2$)
and the cumulant component
\begin{equation}\label{5.2}
{\tilde h}(q,r_s)=\alpha^3\int_0^\infty dx^3\; 
\frac{\sin qx}{qx}h(x,r_s),
\end{equation}
which is simply the Fourier transform of the cumulant pair 
density $h(x,r_s)$. In Eq.~(\ref{5.1}) the convolution theorem
has been applied. $\tilde{n}^2(q,r_s)$ is related to the probability
of finding a pair of electrons with given relative momentum 
$q$.\cite{Gor2,Davo}\par 

Notice that the sum rule 
$S(q\rightarrow 0,r_s)=0$ is equivalent to the sum rule
\begin{equation}\label{5.3}
\alpha^3\int_0^\infty dx^3\; h(x,r_s)=\int_0^\infty dk^3\; 
n(k,r_s)[1-n(k,r_s)]. 
\end{equation}
The l.h.s.~equals ${\tilde h}(0,r_s)$ and the r.h.s.~equals 
$1-\frac{1}{2}{\tilde n}^2(0,r_s)$. 
It was already P.-O. L\"owdin who has asked what meaning the 
r.h.s.~has .\cite{Low} 
According to Eq.~(\ref{5.3}), it fixes the
normalization of the cumulant 
pair density $h(x,r_s)$ and is another particle-hole symmetric
measure of the correlation strength; it is also reported
in Fig.~\ref{fig_corrent}. 
Eq.~(\ref{5.3}) is sometimes called perfect screening sum rule
or charge neutrality condition.\par  

For non-interacting electrons ($r_s=0$), the cumulant 
part vanishes, ${\tilde h}(q,r_s)=0$, and the Fock part 
$S_{\rm F}(q,r_s)$ simply yields
\begin{equation}\label{5.4}
S_0(q)={q\over 2}
\left [{3\over 2}-{1\over 2}\left({q\over 2}\right)^2\right]
\theta \left(1-{q\over 2}\right)+\theta \left({q\over 2}-1\right)
\end{equation} 
with the linear small-$q$ behavior $3q/4$.
For interacting electrons ($r_s\neq 0$) the small-$q$ sum rule
\cite{Pin,Iwa} 
\begin{equation}\label{5.5a}
S(q\ll 1,r_s)=\frac{1}{2(\alpha r_s)^2 
\omega_{\rm pl}(r_s)}\cdot q^2+O(q^4),
\end{equation} 
and the large-$q$ sum rule \cite{Kim3}
\begin{equation}\label{5.5b}
S(q\gg 1,r_s)=1-{8\over {3\pi}}\alpha r_s g_0(r_s)
\cdot \frac{1}{q^4}+O\left(\frac{1}{q^6}\right)
\end{equation}
hold. $\omega_{\rm pl}^2(r_s)=
4\pi e^2\rho/m=3/r_s^3$~a.u.~defines the plasma 
frequency. 

\begin{figure}
\includegraphics[width=7cm]{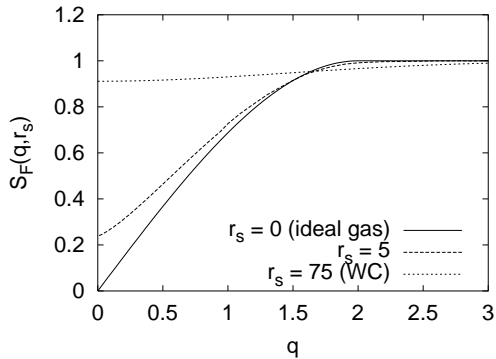}
\caption{The Fock component of the static structure factor for
$r_s=0$ (ideal Fermi gas), $r_s=5$ (present work), and in the
Wigner crystal (WC) limit for $r_s=75$.}
\label{fig_SF}
\end{figure}

The non-idempotency and the singularities of $n(k,r_s)$ 
determine the small-$q$ behavior of $S_{\rm F}(q,r_s)=1-
\frac{1}{2}\tilde{n}^2(q,r_s)$,
\begin{eqnarray}
S_{\rm F}(q \ll 1,r_s)& = & S_{\rm F}(0,r_s)+
\frac{3}{4}z_{\rm F}^2(r_s)q-A(r_s)z_{\rm F}(r_s) \times \nonumber \\
& & q^2 \ln q +
O(q^2),\label{5.5c}
\end{eqnarray}
as shown in Appendix~\ref{app_smallq}. Notice that
$S_{\rm F}(0,r_s)$ is equal to the r.h.s.~of Eq.~(\ref{5.3}).
In Fig.~\ref{fig_SF}, we report
$S_{\rm F}(q,r_s)$ for the ideal gas ($r_s=0$), for
$r_s=5$, and in the WC limit.\par
Eq.~(\ref{5.5c}), together with the 
sum rule of Eq.~(\ref{5.5a}), allows to extract the large-$x$ behavior of 
$h(x,r_s)$. Namely, because of Eq.~(\ref{5.5a}), 
${\tilde h}(q,r_s)$ must cancel both the linear and the 
$q^2\ln q$ term of ${\tilde n}^2(q,r_s)$. This implies that the
large-$x$ behavior of the oscillation-averaged 
$\langle h(q,r_s)\rangle$ is:
\begin{equation}\label{5.5d} 
\langle h(x\gg 1,r_s)\rangle =
-\frac{9}{4}\frac{z_{\rm F}^2(r_s)}{x^4}-
\frac{9}{2}\frac{\pi A(r_s)z_{\rm F}(r_s)}{x^5}+O\left(\frac{1}{x^6}\right).
\end{equation}    
In Eq.~(\ref{4.9}) these terms  cancel 
with the non-oscillatory long-range part of 
$\frac{1}{2}|f(x,r_s)|^2$. 

If we consider within the uniform electron gas a certain fragment
$\Omega$ (e.g., a sphere of radius $R$) containing on average 
$N_{\Omega}=\Omega/(4\pi r_s^3/3)=(R/r_s)^3$ electrons, and ask 
for the particle-number fluctuation $\Delta N_\Omega$ then the 
answer is\cite{Zie8,Zie9}
\begin{eqnarray}
& \frac{(\Delta N_\Omega)^2}{N_\Omega}  =1-\nonumber \\ 
& \frac{(\alpha r_s)^6}{N_\Omega}
\int_\Omega d^3x_1\; \int_\Omega d^3x_2 \; 
\Biggl [{1\over 2}|f(x,r_s)|^2+h(x,r_s)\Biggr ]=1-
\nonumber \\
&   \frac{(\alpha r_s)^6}{N_\Omega}\frac{3}{8\pi}
\int d^3q \Biggl |\int_\Omega d^3x 
{\rm e}^{{\rm i}\underline q \underline x}\Biggr |^2
\left[\frac{1}{2}\tilde{n}^2(q,r_s)+\tilde{h}(q,r_s)\right], 
\label{5.6} \\
& x=|\underline{x}_1-\underline{x}_2|. \nonumber
\end{eqnarray}
Again one may ask how differently the Fock and the cumulant 
parts contribute to their sum and to the conclusion `correlation
suppresses fluctuations' .\cite{Ful,Zie9} In the case of
a sphere $\Omega=4\pi R^3/3$, the term in the modulus in Eq.~(\ref{5.6}) 
is just $\Omega 3 j_1(qR)/(qR)$, so that the Fock term yields
\begin{eqnarray}
& & \left[\frac{(\Delta N_\Omega)^2}{N_\Omega}\right]_{\rm F}  = 
\nonumber \\ & & 1
-\frac{3}{2}\pi^2\Omega
(\alpha r_s)^9 \int_0^{\infty} dq^3\,\left[\frac{3\,j_1(qR)}{qR}
\right]^2\tilde{n}^2(q,r_s).
\end{eqnarray}
Also, the potential energy $v(r_s)$ consists of a Fock and a 
cumulant part:\cite{Zie1}
\begin{eqnarray}
v(r_s) & = & -\frac{\alpha^2}{r_s}\int_0^{\infty}
dx^3\left[{1\over 2}|f(x,r_s)|^2+h(x,r_s)\right]\frac{1}{x}
\nonumber \\
& = & -\frac{\alpha^2}{r_s}\frac{3}{2}\int_0^{\infty}dq^3
\left[\frac{1}{2}\tilde{n}^2(q,r_s)+\tilde{h}(q,r_s)\right]\frac{1}{q^2}
\end{eqnarray}
(in ryd). The Fock part can also be written as~\cite{Zie1}
\begin{eqnarray}
v_{\rm F}(r_s) &=&-\frac{3}{2\pi\alpha r_s}
\int_0^\infty dk\, n(k,r_s) \int_0^\infty dk'\, n(k',r_s)\times
\nonumber \\
& & kk'\ln\frac{k+k'}{|k-k'|}. 
\label{5.7}
\end{eqnarray}
In lowest order, with $n(k,r_s)\to \theta(1-k)$, Eq.~(\ref{5.7})
yields $-3/(2\pi\,\alpha\, r_s)$.
The logarithmic term of $v(r_s\rightarrow 0)$ arises from 
$v_{\rm C}(r_s)$, not from $v_{\rm F}(r_s)$.\cite{Zie4a}
\section{Summary and outlook}
\label{sec_conc}
In Ref.~\onlinecite{Zie3}, it was shown that
the convex Kulik function $G(x)$, decorated with appropriate prefactors and
with an appropriate inhomogeneous scaling of its argument, reproduces the
momentum distribution $n(k,r_s)$ of the unpolarized uniform electron gas of
density $\rho=3/4\pi r_s^3$ in the metallic-density regime
$r_s\in[1,6]$. The $r_s$-functions $n(0,r_s)$, $n(1^\pm, r_s)$, 
the on-top pair density $g(0,r_s)$, and the kinetic energy $t(r_s)$ 
form the input for such construction. 
In this work, we improved the parametrization of $n(k,r_s)$ via
the Kulik function, and we extended it up to 
$r_s\lesssim 12$, including the high-density regime 
[Eqs.~(\ref{eq_nmin1})--(\ref{eq_xmaggiore}),
and Fig.~\ref{fig_allrs}]. \par

The Fourier transform of $n(k,r_s)$ yields the one-body reduced
density matrix $f(x,r_s)$ (Figs.~\ref{fig_onema} and~\ref{fig_n1f1}),
with large $x$-oscillations arising from the Fermi gap
$\zF(r_s)$ and the Fermi edge 
coefficient $A(r_s)$, the prefactor of the
logarithmic term in $n(k\approx 1,r_s)$, which is included
in our parametrization for the first time (Fig.~\ref{fig_a}). 
Several measures of the correlation strength have been discussed
(Fig.~\ref{fig_corrent}). 
With reliable models for the pair density,
$g(x,r_s)$, the cumulant pair density $h(x,r_s)=1-\frac{1}{2}|f(x,r_s)|^2-
g(x,r_s)$ has been extracted (Fig.~\ref{fig_h}),
as a prestep of its diagonalization in terms of cumulant
geminals (analog with the diagonalization of the pair density in terms of
Overhauser geminals). Future work also includes the generalization to
the partially polarized gas. In this case, with $\zeta=
(N_{\uparrow}-N_{\downarrow})/N$, one has to consider different cases.
For spin polarization $\zeta$ between 0 and 1, two
momentum distributions
are to be described, $n_\uparrow(k,r_s, \zeta)$ for the
spin-up electrons and $n_\downarrow(k,r_s,\zeta)$ for the spin-down
electrons. So far, only the input data $g_0(r_s,\zeta)$~\cite{Gor2} and 
$t(r_s,\zeta)$~\cite{Per1} are available in this more general case.
\par
A small {\tt FORTRAN} subroutine, which numerically evaluates
our parametrized $n(k,r_s)$, is available at
{\tt http://axtnt2.phys.uniroma1.it/PGG/elegas.html}.

\section*{Acknowledgments}
The authors thank Y.~Takada for providing the data for $n(0,r_s)$ and
$n(1^\pm,r_s)$, and J.~Cioslowski and G.~Diener for helpful discussions. 
One author
(P.G.-G.) gratefully aknowledges hospitality at the Max Planck Institut
 f\"ur Physik komplexer Systeme of Dresden (Germany), the other author
(P.Z.) thanks P.~Fulde for supporting this work.

\appendix

\section{Random phase approximation}
\label{app_RPA}

In RPA it is\cite{Dan,Kul} $n(k,r_s)=1-(\alpha r_s/\pi^2)^2 H(k,1)$ for $k<1$
and $(\alpha r_s/\pi^2)^2 H(k,1)$ for $k>1$, where
\begin{widetext}
\begin{eqnarray*}
H(x<1,y)& = & \frac{1}{x}\Biggl \{ \int^{1+x}_{1-x}\frac{dq}{q}
\int_0^\infty du \left [ \frac{\frac{1-x^2}{2q}}{(\frac{1-x^2}{2q})^2+u^2}-\frac{{q \over 2}+x}{({q\over 2}+x)^2+u^2}\right ]\frac{Q(q,u)}{q^2+y
\frac{\alpha r_s}{\pi^2}Q(q,u)} \\ 
& & +\int^{\infty}_{1+x}\frac{dq}{q}
\int_0^\infty du \left [\frac{{{q \over 2}-x}}{({q\over 2}-x)^2+u
^2} -\frac{{q \over 2}+x}{({q\over 2}+x)^2+u^2}\right ]\frac{Q(
q,u)}{q^2+y
\frac{\alpha r_s}{\pi^2}Q(q,u)} \Biggr \},\\
H(x>1,y)& = & {1\over x}\int_{x-1}^{x+1} {dq\over q}\int_0^\infty du 
\left [ \frac{\frac{x^2-1}{2q}}{(\frac{x^2-1}{2q})^2+u^2}
-\frac{x-{q\over 2}}{(x-{q\over 2})^2+u^2}\right ] \frac{Q(q,u)}{q^2+y
\frac{\alpha r_s}{\pi^2}Q(q,u)}
\end{eqnarray*}
and
\begin{displaymath}
Q(q,u)=2\pi\left \{ 1+\frac{1+u^2-{q^2/4}}{2q}
\ln\frac{(1+q/2)^2+u^2}{(1-q/2)^2+u^2}
-u\left [\arctan\frac{1+q/2}{u}+\arctan\frac{1-q/2}{u}\right ]
\right \}.
\end{displaymath}
\end{widetext}
For small or large $k$ (far from the Fermi edge)
it is $H(k,1)\rightarrow F(k)$ with $F(k)=H(k,0)$. 
$F(k)$ has the small and large $k$ properties
\begin{equation}
F(k\ll 1)=4.112335+8.984373 \cdot k^2 +O(k^4)
\label{eq_Fsmallk}
\end{equation}
and
\begin{equation} F(k\gg 1)=\frac{8\pi^2}{9}\cdot \frac{1}{k^8}+
O\left({1\over k^{10}}\right),
\label{eq_Flargeq}
\end{equation}
respectively .\cite{Cio2,Cio3} The coefficient of $1/k^8$ is 
8.77298.
For $k$ near the Fermi edge it is \cite{Cio2,Cio3}
\begin{equation}
H(k,1)\rightarrow \frac{\pi^2}{2\alpha r_s}{1\over k^2}
G\left (\frac{|k-1|}{\sqrt {4\alpha r_s/\pi}}\right )
\label{eq_RPAnear1}
\end{equation}
with
\begin{equation}
G(x)=\int_0^\infty du \,\frac{R'(u)}{R(u)}\cdot 
\frac{u}{u+y}
\cdot \frac{R(u)-R(y)}{u-y}\Biggr |_{y=x/{\sqrt {R(u)}}}
\label{eq_G(x)}
\end{equation} 
and 
\begin{equation}
R(u)=1-u\arctan {1\over u}.
\label{eq_R(u)}
\end{equation}
$G(x)$ has the small-$x$ behavior \cite{Kul}
\begin{equation}
G(x\ll 1)=G(0)+\left[\pi\left(\tfrac{\pi}{4}+\sqrt{3}\right) \cdot x +
O(x^2)\right]\ln x +O(x)
\label{eq_Gedge}
\end{equation}
with 
\begin{equation}
G(0)=\int_0^\infty du\; (-1)\frac{R'(u)}{R(u)}\cdot 
\arctan {1\over u}\approx 3.353337 
\label{eq_Gof0}.
\end{equation}
The coefficient of $x\ln x$ is 7.908799 (the Kulik number). $G(x)$ has the 
large-$x$ behavior
\begin{equation}
G(x\gg 1)=\frac{\pi}{6}(1-\ln 2)\cdot {1\over x^2} +O\left({1\over x^4}\right).
\end{equation} 
The coefficient of $1/x^2$ is 0.160668 (the Macke number). The
Kulik function $G(x)$ is shown in Fig.~\ref{fig_kul}.

\section{The momentum distribution of the Wigner crystal} 
\label{app_Wigner}

In the low-density (large $r_s$) or strongly correlated limit,
the electrons localize~\cite{Ort2} and form a 
ferromagnetic body-centered cubic lattice with an electrostatic (or Madelung)
energy of $-1.792/r_s$~ryd.\cite{Fuc} The next term, 
$+2.65/r_s^{3/2}$~ryd, describes
the coupled harmonic zero-temperature motion in lowest order 
.\cite{Cold,Car1,Car2} To estimate the corresponding $n(k,r_s)$,
we define with $3\hbar \omega/2=2.65/r_s^{3/2}$ (in ryd, or 
$\omega =0.88/r_s^{3/2}$ in a.u.) the frequency of indepent 
oscillating electrons (Einstein model). So, from the momentum 
distribution of the
harmonic-oscillator ground state it follows
\begin{equation}\label{A2.1}
n(k,r_s\rightarrow\infty)=\frac{4\pi}{3}
\frac{1}{(\pi \omega/k_{\rm F}^2)^{3/2}}
{\rm e}^{-\frac{k^2}{\omega/k_{\rm F}^2}}
\end{equation}
see Refs.~\onlinecite{Die,Mar2}, and~\onlinecite{Zie5}, p.~19. 
Note that $k$ is 
dimensionless (measured in units of $k_{\rm F}$) and that 
$\omega/k_{\rm F}^2=0.88\cdot \alpha^2r_s^{1/2}=0.24 \cdot r_s^{1/2}$. 
In Ref.~\onlinecite{Mar2}, the
factor 1 is used instead of 0.88. $n(k,r_s\rightarrow\infty)$ is
correctly normalized and yields with Eq.~(\ref{eq_kinetic}) the kinetic 
energy (in ryd)
\begin{equation}\label{A2.2}
t(r_s\rightarrow\infty)={1\over 2}\cdot \frac{2.65}{r_s^{3/2}}+\cdots
\end{equation}
as it should. The corresponding potential energy is 
\begin{equation}\label{A2.3}
v(r_s\rightarrow\infty)=-\frac{1.792}{r_s}+{1\over 2}\cdot
\frac{2.65}{r_s^{3/2}}+\cdots .
\end{equation}
The inflexion-point trajectory with $r_s$ as parameter is
described by (see left panel of Fig.~\ref{fig_zF})
\begin{eqnarray}
k_{\rm infl}(r_s) & = & (\omega/(2k_{\rm F}^2))^{1/2}=0.35 \cdot r_s^{1/4},
 \nonumber \\ 
n_{\rm infl}(r_s) & = &
\frac{4\pi}{3}
\frac{e^{-1/2}}{(\pi \omega/k_{\rm F}^2)^{3/2}}
 =  \frac{3.88}{r_s^{3/4}}.
\label{A2.4}
\end{eqnarray} 
The region $k>k_{\rm infl}$ (maybe to be referred to as 
correlation tail) contributes to the normalization the constant
amount
\begin{equation}
\int_{k_{\rm infl}}^\infty dk^3\; n(k,r_s)= 
{\rm Erf}({1\over \sqrt 2})-\sqrt {2\over {\rm e} \pi}=0.80. 
\end{equation}
From Eq.~(\ref{A2.1}) it follows for $n(0,r_s\rightarrow\infty)$
\begin{equation}\label{A2.5}
n_0(r_s\rightarrow\infty)=\frac{4}{3\pi^{1/2}}
\Biggl (\frac{1}{0.24 r_s^{1/2}}\Biggr )^{3/2}=\frac{6.40}{r_s^{3/4}},
\end{equation}
see Fig.~\ref{fig_n0};
and for the curvature at the centre (the coefficient of $k^2$)
\begin{equation}\label{A2.6}
-\frac{4}{3\pi^{1/2}}\frac{1}{(0.88\alpha^2r_s^{1/2})^{5/2}}=-
\frac{26.71}{r_s^{5/4}},
\end{equation}
see Fig.~\ref{fig_b}.\par
A more refined treatment takes into account that in harmonic 
approximation there are two transversal branches of harmonic 
lattice vibrations, $\omega_{t_{1,2}}(\underline q,r_s)$, and 
one longitudinal branch $\omega_l(\underline q,r_s)$ 
in the face-centred cubic Brillouin zone, satisfying
the sum rule $\omega_{t_{1}}^2(\underline q,r_s)+
\omega_{t_{2}}^2(\underline q,r_s)+
\omega_l^2(\underline q,r_s)=\omega_{\rm pl}^2(r_s)$. For 
$q=0$ it is $\omega_{t_{1,2}}(0,r_s)=0$ and therefore 
$\omega_l(0,r_s)=\omega_{\rm pl}(r_s)$. But also in this case 
the virial theorem holds and assuming that $n(k,r_s)$ is a 
Gau\ss \ distribution, then Eq.~(\ref{A2.1}) turns out again.  
For phonons in Wigner crystals near melting see
also Ref.~\onlinecite{tosi}.

\section{1-matrix near the diagonal and far from it}
\label{app_diag}

The equation $\gamma(1|1')=\rho\,\delta_{\sigma_1\sigma_1'}\,f(\kF |
{\underline r}_1-{\underline r}_1'|)$ defines the dimensionless 1-matrix
$f(x)$. Its
small-$x$ behavior of Eq.~(\ref{4.6}) follows from the large-$k$
behavior of $n(k)$ [Eq.~(\ref{eq_nlargek})]. Namely, with $\sin y/y=1-y^2/3!+
y^4/5!-\cdots$ and with the integrability of $n(k)k^2k^\nu$ for 
$\nu=0,\dots, 4$ it is
\begin{equation}
f^{(\nu)}(0)= \int^\infty_0 d k^3 \; n(k)k^\nu\left(\frac{d}{dy}\right)^\nu
\frac{\sin y}{y}\Biggr |_{y=0}\; ,
\end{equation} 
which yields the first three terms of Eq.~(\ref{4.6}). 
Here $f^{(\nu)}(0)=(\frac{d}{dx})^{\nu}f(x)|_{x=0}$. \par

Because $n(k)k^2k^5$ is non-integrable, one has to compute 
$f^{(5)}(0)$ with the Kimball procedure,\cite{Kim3,Raj} which 
defines by
\begin{equation}
n(k)=\frac{C}{(1+k^2)^4}+\mathcal{N}(k), \quad \quad C=\frac{8}{9 \pi^2}
(\alpha r_s)^2 g_0(r_s)
\end{equation}
a stronger (namely $\sim 1/k^{10}$ for $k\rightarrow \infty)$ 
decaying function $\mathcal{N}(k)$, so that $\mathcal{N}(k)k^2k^5$
is now integrable, yielding 0 because of 
$(d/dy)^5\sin y/y|_{y=0}=0$. Thus 
\begin{equation}
f^{(5)}(0)=\Biggl (\frac{d}{dy}\Biggr )^5 p(x)\Biggr |_{x=0}
\end{equation}
with 
\begin{equation}
p(x)=C\int^\infty_0\frac{dk^3}{(1+k^2)^4}\frac{\sin kx}{kx}=
C\frac{\pi}{32}(3+3x+x^2){\rm e}^{-x} .
\end{equation}
It follows
\begin{equation}
f^{(5)}(0)=-C\frac{\pi}{4}=-\frac{2}{9\pi}
(\alpha r_s)^2 g_0(r_s),
\end{equation}
q.e.d. \par

The large-$x$ behavior (\ref{4.7}) follows from Eq.~(\ref{4.5})
by partial integration. Thereby the discontinuities of 
$n(k,r_s)$ at $k\approx 1$ appear. They determine the amplitudes 
of the Friedel oscillations:
\begin{equation}
f(x,r_s)=-z_{\rm F}(r_s)\frac{3\cos x}{x^2}+
z_{\rm F}(r_s)\frac{3\sin x}{x^3}+f_1(x,r_s)
\end{equation}
with
\begin{eqnarray}
f_1(x\gg 1,r_s) & = &
-A(r_s)\frac{3}{x^3}\{[{\pi\over 2}+{\rm Si}(x)]\cos x -
{\rm Ci}(x) \times \nonumber \\ 
& & \sin x \} +O\left({1\over x^4}\right)
\nonumber \\
& = & -A(r_s)\pi \frac{3 \cos x}{x^3}
+O\left({1\over x^4}\right).
\end{eqnarray}   
$z_{\rm F}(r_s)$ is the Fermi gap and $A(r_s)$ is the Fermi edge
coefficient. 

\section{Fock component of the static structure factor at small~$q$}
\label{app_smallq}

According to the definition of Eq.~(\ref{5.1}), the oscillations of 
$|f(x\gg 1,r_s)|^2$ [see Eq.~(\ref{4.11})] only affect the 
discontinuities in ${\tilde n}^2(q,r_s)$ and in its 
derivatives at $q=2$, while the small-$q$ behavior of 
${\tilde n}^2(q,r_s)$ is only affected by the 
oscillation-averaged part of $|f(x,r_s)|^2$, i.e., by
\begin{equation}
\langle f^2(x\gg 1,r_s)\rangle 
=\frac{9}{2}\frac{z_{\rm F}^2(r_s)}{x^4}
+9\frac{\pi A(r_s)z_{\rm F}(r_s)}{x^5}+
O\left(\frac{1}{x^6}\right).
\end{equation} 
Following the procedure of Kimball \cite{Kim3,Raj} we define a function
$\mathcal{F}(x)$ by
\begin{equation}
\langle f^2(x,r_s)\rangle =\mathcal{F}(x)+\frac{9}{2}
\frac{z_{\rm F}^2(r_s)}{(1+x^2)^2}+
9\frac{\pi A(r_s)z_{\rm F}(r_s)}{(1+x)^5},
\end{equation}
so that 
$\mathcal{F}(x\rightarrow \infty)\propto1/x^6$.
Then, the second term will give the coefficient of the linear 
term in ${\tilde n}^2(q,r_s)$, while the third term will give 
the coefficient of a term $\propto q^2 \ln q$. By carrying out 
the calculations one obtains Eq.~(\ref{5.5c}), q.e.d. .


\begin{thebibliography}{99}
\bibitem{notaSav}   The infinite slope at the Fermi edge was previously 
taken into account in a study
of the momentum distribution of $^3$He;\cite{Mor}
the authors also extracted the Fermi edge coefficient $A$ of
Eqs.~(\ref{eq_lim1-}) and~(\ref{eq_lim1+}) from QMC simulations.
At the equilibrium density they found $A=0.06(2)$.
\bibitem{notaRPA} Comparison of Eq.~(\ref{eq_Flargeq}) with
Eq.~(\ref{eq_nlargek}) shows that RPA fails to reproduce
the correct large-$k$ expansion of $n(k,r_s)$: for small-$r_s$,
instead of having $k^8n(k)\to\frac{8}{9}\pi^2(\alpha r_s^2)[\frac{1}{2}
+O(r_s)]$, RPA yields $k^8n(k)\to\frac{8}{9}\pi^2(\alpha r_s^2)[1+
O(r_s)]$. The lowest order exchange diagram~\cite{Ons} repairs this,
see Figs.~1c and~1f of Ref.~\onlinecite{Zie4a}.
\end{thebibliography}
\end{document}